\documentclass[11pt]{article}

\usepackage[utf8]{inputenc}
\usepackage[T1]{fontenc}
\usepackage{lmodern}
\usepackage{geometry}
\usepackage{microtype}
\usepackage[numbers,sort&compress]{natbib}
\usepackage{hyperref}
\usepackage{booktabs}
\usepackage{tabularx}
\usepackage{graphicx}
\usepackage{amsmath,amssymb}
\usepackage{enumitem}

\hypersetup{
  colorlinks=true,
  linkcolor=blue,
  citecolor=blue,
  urlcolor=blue
}

\title{Self-hosted Lecture-to-Quiz: Local LLM MCQ Generation with Deterministic Quality Control}

\author{
Seine A.\ Shintani\\
Department of Biomedical Science, College of Life and Health Sciences, Chubu University\\
Center for AI, Mathematical and Data Sciences, Chubu University\\
Kasugai, Aichi 487-8501, Japan\\
ORCID: 0000-0002-1084-2549
}

\date{February 2026}

\begin{document}
\maketitle

\begin{abstract}
We present an end-to-end \emph{self-hosted} (API-free) pipeline---API-free in the sense that lecture content is \emph{not} sent to any external LLM service---that converts lecture PDFs into multiple-choice questions (MCQs) using a local LLM plus deterministic quality control (QC).
The pipeline is designed for \emph{black-box minimization}: LLMs may assist drafting, but the final released artifacts are plain-text question banks with an explicit QC trace and without any need to call an LLM at deployment time.
We run a seed sweep on three short ``dummy lectures'' (information theory, thermodynamics, and statistical mechanics), collecting 15 runs $\times$ 8 questions = 120 accepted candidates (122 attempts total under bounded retries).
All 120 accepted candidates satisfy hard QC checks (JSON schema conformance, a single marked correct option, and numeric/constant equivalence tests); however, the warning layer flags 8/120 items (spanning 8 runs) that expose residual quality risks such as duplicated distractors or missing rounding instructions.
We report a warning taxonomy with concrete before$\rightarrow$after fixes, and we release the final 24-question set (three lectures $\times$ 8 questions) as JSONL/CSV for Google Forms import (e.g., via Apps Script / API tooling) (included as ancillary files under \texttt{anc/}).
Finally, we position the work through the AI to Learn (AI2L) rubric lens and argue that self-hosted MCQ generation with explicit QC supports privacy, accountability, and Green AI in educational workflows.
\end{abstract}

\section{Introduction}
Large language models (LLMs) can draft explanations, examples, and questions that support learning.
For assessment authoring, however, instructors often face constraints that make ``prompt-and-publish'' workflows unsafe: materials may be proprietary, the correctness of generated items must be auditable, and repeated use of large models can be costly in both time and energy.

This paper focuses on one practical scenario: converting lecture PDFs into multiple-choice questions (MCQs) in a \emph{self-hosted} (API-free) manner, meaning generation that does \emph{not} transmit lecture content to external LLM services. Throughout, we use this narrow API-free definition; data residency depends on the execution environment (e.g., fully local vs.\ hosted notebooks).
Using a self-hosted model reduces privacy risk, but local generation raises its own challenges: outputs can contain duplicated options, multiple correct answers, or numerically equivalent distractors that silently break the question.

We present \emph{Self-hosted Lecture-to-Quiz (L2Q)}, a local-LLM pipeline with an explicit verification layer.
L2Q (i) extracts a compact topic plan from the lecture, (ii) generates MCQs in a strict JSON schema, and (iii) applies deterministic quality-control (QC) checks (duplicate detection, numeric/constant and sampled equivalence tests, and unique-correct verification) with bounded retries.
The end product is a static question set that can be imported into systems such as Google Forms; learners do \emph{not} need access to an LLM at deployment time.

\paragraph{Positioning with AI2L (AI to Learn).}
AI2L is a set of human-centered guidelines that constrain generative AI to the \emph{learning-support stage} and aim to eliminate AI-derived black-box components from final deliverables, emphasizing (1) model transparency, (2) accountability via human verification, (3) privacy protection, and (4) Green AI through ``use big models only when needed'' and deploy lightweight artifacts \cite{shintani2025ai2l}.
Our study is \emph{not} a direct application of AI2L in the strict sense, because MCQ drafting still relies on a black-box LLM.
Instead, we treat L2Q as a concrete case study of \emph{black-box minimization}: we run locally to address privacy (pillar~3), we provide QC logs and a rubric-friendly workflow to support human accountability (pillar~2), and we export static quizzes that no longer require repeated LLM calls during learning activities (pillar~4).
We discuss what is achieved and what remains unresolved with respect to pillar~1 (full transparency) in the Discussion (Table~\ref{tab:ai2l_lens}).

\paragraph{Contributions.}
\begin{itemize}
  \item We propose an end-to-end self-hosted (API-free) pipeline (L2Q) for converting lecture PDFs into deployable MCQs using a local LLM.
  \item We integrate deterministic QC checks (duplicate detection, equivalence testing, unique-correct verification) with bounded retries to reduce structural failure modes.
  \item We report a 15-run case study on an entropy lecture pack, producing a curated set of 24 deployable MCQs and summarizing runtime and QC statistics.
  \item We position the workflow with respect to the AI2L guidelines, clarifying both the AI2L-compatible aspects (privacy, accountability, reduced runtime dependence) and the remaining black-box limitations.
\end{itemize}
\section{Related Work}
\paragraph{Automatic question generation.}
Automatic question generation (AQG) has been studied for decades, ranging from rule- and template-based methods to neural approaches \cite{heilman2010good,du2017learning}. Benchmark datasets such as RACE \cite{lai2017race} and SQuAD \cite{rajpurkar2016squad} helped standardize evaluation, although classroom deployment typically requires additional constraints (format stability, answer uniqueness, and pedagogical review).

\paragraph{LLM-based MCQ generation workflows.}
With instruction-tuned LLMs, MCQ generation is increasingly treated as a workflow problem: prompting strategies, multi-stage generation, and filtering are combined with human review to reach deployable quality. Recent studies explore multi-stage prompting for MCQs \cite{maity2024multistage}, reliability analyses and educator perspectives \cite{biancini2025educator}, and cognitively diverse MCQ generation with psychometric-style evaluation \cite{tian2026requesta}. Education-specific deployments also emphasize teacher revision loops (e.g., stepwise MCQ generation for English education) \cite{wang2024aqg}. These works mainly optimize question quality given access to powerful models (often cloud-hosted). Our focus is complementary: we emphasize \emph{self-hosted (API-free)} generation and deterministic QC gates, aiming to reduce privacy risk and to avoid a permanent black-box dependency during learning sessions.

\paragraph{Structured outputs and verification.}
Reliable downstream use benefits from constrained decoding and post-hoc verification. Constrained decoding has a substantial prior literature (e.g., lexically constrained decoding) that improves format and content controllability during generation \cite{post2018fast}. We employ \texttt{llama.cpp} grammar-constrained decoding to stabilize JSON outputs \cite{llamacpp} and add deterministic checks for uniqueness and numeric/constant equivalence. This supports black-box minimization by turning model outputs into auditable artifacts, while acknowledging that the drafting model itself remains opaque.

\section{Method: Self-hosted Lecture-to-Quiz (L2Q) Pipeline}
L2Q is designed as a deterministic \emph{outer loop} around a generative model.
The pipeline has five stages:
\begin{enumerate}[leftmargin=*]
\item \textbf{PDF ingestion and segmentation.} Extract text from a lecture PDF and split into coherent chunks with page references.
\item \textbf{Topic planning.} Produce a short topic plan (e.g., key definitions and properties) to encourage coverage and reduce redundancy.
\item \textbf{MCQ drafting.} Generate MCQs with five options (A--E), a single correct option, and a brief explanation.
\item \textbf{Automatic QC and bounded retries.} Apply a suite of tests; if a test fails, re-prompt the model with targeted feedback up to a fixed retry budget.
\item \textbf{Export.} Emit JSONL/CSV suitable for LMS ingestion and Google Forms import (e.g., via Apps Script / API tooling).
\end{enumerate}

\subsection{Quality-control tests}
We implement QC as a two-level rule set: \emph{hard} constraints (reject + retry) and \emph{warning} flags (accept but log for human review). The hard checks target machine-verifiable failures that break MCQs in practice; warning checks capture ``editorial'' risks that are inexpensive for humans to fix.

\paragraph{Hard QC (reject + retry).}
\begin{itemize}[leftmargin=*]
\item \textbf{Schema compliance.} Output must be valid JSON with required keys and exactly five options (A--E).
\item \textbf{Single-answer structure.} Exactly one labeled correct option; the correct option must be in \{A,\dots,E\}.
\item \textbf{De-duplication.} Reject exact duplicates and near-duplicates (string similarity $\ge 0.92$) within a run.
\item \textbf{Equivalence and uniqueness.} For options that are numeric expressions, evaluate constant equivalence within tolerance $10^{-9}$; for parametric expressions, test equivalence across 5 random trials. Reject if any option is equivalent to the labeled correct answer (unique-correct verification).
\end{itemize}

\paragraph{Warning QC (accept + log).}
\begin{itemize}[leftmargin=*]
\item \textbf{Rounding instruction missing (Rnd).} If the correct answer is a non-integer numeric approximation (e.g., $0.881$ bits) but the stem does not specify rounding, flag for minor revision.
\item \textbf{Duplicate constant distractors (Dup).} If two distractors evaluate to the same constant, flag for re-generation or human editing.
\end{itemize}

When a hard check fails, we re-sample the item up to \texttt{retry\_max} times and record all attempts in the run manifest.

\section{Implementation and reproducibility}
We implement L2Q in Python as a deterministic wrapper around \texttt{llama.cpp} \cite{llamacpp}.
We provide a reference implementation on Google Colab for ease of re-running the experiments; however, the pipeline is platform-agnostic and can also run on a local PC (GPU recommended; CPU is possible with slower throughput) because it depends only on local inference via \texttt{llama.cpp} and GGUF weights.
Importantly, under our definition, using Colab does not change the API-free property as long as no lecture content is sent to an external LLM API. However, API-free is distinct from on-premises data residency: running on Colab uploads PDFs to a hosted runtime; institutions with strict data-residency requirements should run the same pipeline on local hardware.
A \emph{reproducible run} is defined by the tuple $(\texttt{PDF},\ \texttt{seed},\ \texttt{model\_spec},\ \texttt{QC\_spec})$; these components are logged to support re-running under the same recorded setup. In practice, minor differences can still occur across toolchain versions, build flags, and hardware, so we treat manifests and checksums as the primary reproducibility contract.

\begin{table}[t]
\centering
\caption{Reproducible run configuration (shared across all runs unless noted).}
\label{tab:config}
\begin{tabular}{lp{0.66\linewidth}}
\toprule
Component & Setting \\
\midrule
Execution & Google Colab Pro+ (NVIDIA A100) \\
Model & Qwen2.5-14B-Instruct (GGUF, Q4\_K\_M) via \texttt{llama-cli} \cite{qwen2techreport,qwen25hf,bartowskiqwen25gguf,llamacpp} \\
Context window & 4096 \\
Decoding & temperature 0.2, top\_p 0.9, max\_new\_tokens 900 \\
Seeds & \{0,1,2,3,4\} (seed sweep; recorded per run) \\
Retry budget & \texttt{retry\_max} = 3 per question \\
Duplicate checks & exact match + fuzzy similarity threshold 0.92 \\
Equivalence checks & const tolerance $10^{-9}$; parametric trials 5; unique-correct enabled \\
Outputs & run-level JSON, QC report, JSONL/CSV exports \\
\bottomrule
\end{tabular}
\end{table}

\paragraph{Prompt template and output contract.}
We use a fixed question-generation prompt that (i) restricts content to the provided lecture excerpt, (ii) requests exactly \emph{one} MCQ per model call and repeats the call with different topic hints until 8 accepted items are collected per run, (iii) enforces five options (A--E) and a single correct answer, and (iv) requires strict JSON output with keys \{\texttt{SourcePages}, \texttt{Question}, \texttt{OptionA}--\texttt{OptionE}, \texttt{CorrectOption}, \texttt{Explanation}, \texttt{ID}, \texttt{Category}, \texttt{FocusTopic}\}.
The full prompt template and run artifacts are included as ancillary files under \texttt{anc/} and summarized in Appendix~\ref{app:prompt}.

\paragraph{Logging and exports.}
For each run, we save \texttt{run\_manifest.json} with the configuration, timing, and file checksums (and, where available, tool/version identifiers such as the \texttt{llama.cpp} commit and GGUF hash), together with run-level artifacts (\texttt{questions.json} and \texttt{QuestionBank.csv}) and a QC issues log (\texttt{auto\_check\_issues.json}) recording warning flags.
For the curated deploy set, we additionally provide normalized exports in JSONL and CSV for use in quiz systems (e.g., Google Forms via Apps Script / API tooling).

\section{Experimental Setup}
\subsection{Data}
We evaluate on three short lecture notes about entropy (2--3 pages each), covering:
(i) information-theoretic entropy, (ii) thermodynamics and the second law, and (iii) statistical-mechanical entropy.
The notes are treated as self-contained inputs for lecture-to-quiz conversion.

\subsection{Protocol and metrics}
For each lecture we run the full pipeline with five random seeds.
Each run targets 8 MCQs, yielding 120 questions in total.
We report:
(i) end-to-end runtime per run, (ii) QC acceptance and warning counts, and (iii) a qualitative summary of flagged issue types.
Finally, we select one seed per lecture to form a 24-question deployable set.

\section{Results}
\noindent\textbf{We demonstrate three practical outcomes.}
\begin{enumerate}[leftmargin=*,itemsep=0.2em]
\item \textbf{Seed-sweep stability and cost:} 15 runs (3 lectures $\times$ 5 seeds) produced 120 draft MCQs; only 2 additional generation attempts were needed (122 attempts total; retry rate 2/122 = 1.6\%).
\item \textbf{Quality-risk visibility via warnings:} the hard QC gate accepted 120/120 items (1.0), while the warning system flagged 8/120 items (6.7\%) for targeted human polishing, dominated by rounding-instruction omissions.
\item \textbf{Reviewable, deployable artifacts:} QC traces support human review and \emph{black-box minimization}; we curated a 24-item deploy set with zero warnings and provide JSONL/CSV exports for Google Forms.
\end{enumerate}
Table~\ref{tab:seed_sweep} summarizes per-run metrics; Table~\ref{tab:agg_perf} aggregates runtime, retry, and warning rates.

\subsection{Seed sweep stability and quality-risk visibility}
Across three lectures and five random seeds each, runtimes were stable (Table~\ref{tab:agg_perf}) and retries were rare. More importantly, QC warnings made latent quality risks \emph{observable} without requiring any additional experiments: Table~\ref{tab:seed_sweep} shows which runs emitted which warning flags, and Table~\ref{tab:issue_breakdown} summarizes warning types and counts.

\begin{table}[t]
\centering
\caption{Aggregate performance over the 15-run seed sweep (3 lectures $\times$ 5 seeds; 8 items target per run).}
\label{tab:agg_perf}
\begin{tabular}{lrr}
\toprule
Metric & Value & Notes \\
\midrule
Runs & 15 & 3 lectures $\times$ 5 seeds \\
Target items & 120 & 8 per run \\
QC acceptance & 120/120 & Acceptance rate 1.0 \\
Total attempts & 122 & Includes retries \\
Retry rate & 1.6\% & 2/122 attempts (1.7\% overhead per item) \\
Runtime / run & 58.55 $\pm$ 2.34 s & Min 55.29, max 63.50 \\
Runtime / item & 7.32 $\pm$ 0.29 s & 8 items/run \\
Warning flags & 8 (6.7\%) & Non-fatal; review recommended \\
Final deploy set & 24 & 3 selected runs (0 warning flags) \\
\bottomrule
\end{tabular}
\end{table}

\begin{table}[t]
\centering
\small
\caption{Seed sweep results (target 8 items per run). Retries indicate extra attempts beyond 8 triggered by hard QC failures. Flags are warning-level checks: Rnd = missing rounding instruction, Dup = duplicated numeric constant distractors.}
\label{tab:seed_sweep}
\resizebox{\linewidth}{!}{
\begin{tabular}{llrrrrrl}
\toprule
Lecture & Seed & Runtime (s) & s/q & Tries & Retries & Retry (\%) & Flags \\
\midrule
InfoTheory & 0 & 57.22 & 7.15 & 8 & 0 & 0.0 & Dup \\
InfoTheory & 1 & 57.02 & 7.13 & 8 & 0 & 0.0 & Rnd \\
InfoTheory & 2 & 57.81 & 7.23 & 8 & 0 & 0.0 & -- \\
InfoTheory & 3 & 56.86 & 7.11 & 8 & 0 & 0.0 & -- \\
InfoTheory & 4 & 63.50 & 7.94 & 9 & 1 & 11.1 & -- \\
Thermodynamics & 0 & 58.84 & 7.35 & 8 & 0 & 0.0 & -- \\
Thermodynamics & 1 & 56.79 & 7.10 & 8 & 0 & 0.0 & Rnd \\
Thermodynamics & 2 & 57.88 & 7.24 & 8 & 0 & 0.0 & Rnd \\
Thermodynamics & 3 & 58.76 & 7.35 & 8 & 0 & 0.0 & Rnd \\
Thermodynamics & 4 & 58.80 & 7.35 & 8 & 0 & 0.0 & Rnd \\
StatMech & 0 & 59.15 & 7.39 & 8 & 0 & 0.0 & -- \\
StatMech & 1 & 56.39 & 7.05 & 8 & 0 & 0.0 & Rnd \\
StatMech & 2 & 62.82 & 7.85 & 9 & 1 & 11.1 & Rnd \\
StatMech & 3 & 55.29 & 6.91 & 8 & 0 & 0.0 & -- \\
StatMech & 4 & 61.16 & 7.64 & 8 & 0 & 0.0 & -- \\
\bottomrule
\end{tabular}}
\end{table}

\begin{table}[t]
\centering
\small
\caption{Warning-level issue breakdown across the seed sweep (120 accepted items).}
\label{tab:issue_breakdown}
\begin{tabular}{lrrp{0.56\linewidth}}
\toprule
Flag & Count & Share & Typical cause and mitigation \\
\midrule
Rnd & 7 & 5.8\% & Numeric approximation without explicit rounding instruction; mitigate by adding ``approximately'' or specifying rounding digits in the stem/template. \\
Dup & 1 & 0.8\% & Two distractors evaluate to the same constant; mitigate via equivalence checks + re-generation or by constraining distractor generation. \\
\bottomrule
\end{tabular}
\end{table}

\subsection{QC warning fixes (before $\rightarrow$ after)}
QC warnings are intended to be \emph{actionable}: they point to concrete, fixable issues that are easy to miss when skimming a large question bank.
Table~\ref{tab:qc_fixes} illustrates minimal revisions for the two warning types observed in our sweep.
These examples come from seed runs that were \emph{not} used for the final deploy set; we include them to make the ``quality risks that would otherwise slip through'' visible without introducing any new experiments.

\begin{table}[t]
\centering
\caption{Representative QC warning fixes (before$\rightarrow$after).}
\label{tab:qc_fixes}
\begin{tabularx}{\linewidth}{l X X}
\toprule
Warning type & Before (flagged excerpt) & After (minimal fix) \\
\midrule
duplicate\_options\_numeric\_const &
\textbf{L01\_Q01 (seed 0)} included two options that evaluate to the same constant: \textbf{C} $\log_2(0.25)=-2$ and \textbf{E} $2\cdot \log_2(0.5)=-2$. &
Replace one distractor with a non-equivalent expression (e.g., \textbf{E} $\log_2(0.125)=-3$) so that all five options are distinct. \\
rounding\_instruction\_missing\_suspect &
\textbf{L01\_Q02 (seed 1)} asked for $H(X)$ for $(0.7,0.3)$ and offered a decimal correct option (0.881 bits) without an explicit rounding rule. &
Add a rounding contract in the stem (e.g., ``Round to three decimal places'') \emph{or} redesign the item to ask for the symbolic expression $-0.7\log_2 0.7-0.3\log_2 0.3$. \\
\bottomrule
\end{tabularx}
\end{table}

\begin{table}[t]
\centering
\small
\begin{tabular}{lrrrrr}
\toprule
Set & Def./formula & Property/law & Worked ex. & Mean stem (words) & Mean expl. (words)\\
\midrule
L01 & 2 & 5 & 1 & 18.4 & 37.8\\
L02 & 4 & 4 & 0 & 22.5 & 48.4\\
L03 & 5 & 3 & 0 & 28.4 & 43.2\\
Overall & 11 & 12 & 1 & 23.1 & 43.1\\
\bottomrule
\end{tabular}
\caption{A lightweight profile of the final 24-item deploy set. ``Def./formula'' includes definition- and equation-centric items; ``Property/law'' includes qualitative statements and inequalities; word counts are computed over the English text fields in the released JSONL/CSV.}
\label{tab:dataset_profile}
\end{table}

\subsection{Representative outputs: what a ``good'' item looks like}
Beyond aggregate statistics, the output of our pipeline is a self-contained MCQ item with (i) an explicit source-page pointer, (ii) five options with a \emph{single} correct answer, and (iii) a short explanation intended to make the answer auditable by a human reviewer and useful as immediate feedback for learners.
As a concrete illustration, the following item is taken verbatim from the final deployment set (the appendix lists additional examples).

\begin{quote}\small
\textbf{Example (L01\_Q07; p.1--3; biased coin entropy).}
\emph{Question:} For a biased coin with $p(1)=0.9$ and $p(0)=0.1$, what is the approximate Shannon entropy $H(X)$ in bits?\\
\emph{Options:} (A) 0.1368 bits; (B) 0.3322 bits; (C) 0.469 bits; (D) 0.6 bits; (E) 1 bit.\\
\emph{Correct:} C. \emph{Rationale:} $H(X)=-0.9\log_2(0.9)-0.1\log_2(0.1)\approx 0.469$ bits.
\end{quote}

Even in this short example, the design intent is visible: distractors are numerically plausible (values that could arise for other biases), while the explanation makes the computation transparent enough for quick human verification.
For a broader view of the writing style and option structure across lectures, see Appendix~\ref{app:sample_mcqs}.

\subsection{Worked QC trace: what the warnings mean in practice}
The QC layer is intentionally conservative: it is designed to surface \emph{potential} ambiguities early, even at the cost of occasional false positives.
Table~\ref{tab:qc_case_studies} shows representative warning traces from the seed sweep and how they are resolved.

\begin{table}[t]
\centering
\small
\begin{tabularx}{\linewidth}{l l l X}
\toprule
Run & Item & Warning & Why it was flagged / how it is handled \\
\midrule
L01 (seed 0) & Q01 & duplicate\_options\_numeric\_const &
Two distractors were constant-equivalent: \texttt{log2(0.25)} and \texttt{2*log2(0.5)} both evaluate to $-2$. We resolve this either by regenerating a single distractor or by selecting another seed run without the duplication. \\
L01 (seed 1) & Q02 & rounding\_instruction\_missing\_suspect &
The keyed answer was reported as \texttt{0.881 bits} but the stem did not specify whether rounding (and to how many digits) was required. A minimal fix is to add ``approximately'' or an explicit rounding rule. \\
L02 (seed 1) & Q06 & rounding\_instruction\_missing\_suspect &
A conservative heuristic was triggered by a comparison to a numerical constant (``$\ge 0$'') despite the item being qualitative. This illustrates why the warnings are best interpreted as a \emph{review queue}, not as automatic rejection. \\
\bottomrule
\end{tabularx}
\caption{Representative QC warning traces observed in the seed sweep. In deployment, we typically resolve warnings by (i) regenerating only the problematic component or (ii) applying a small human edit after verifying correctness.}
\label{tab:qc_case_studies}
\end{table}

\subsection{From a question bank to practice: export and deployment}
After QC and (optional) human review, the question bank is exported to JSONL/CSV.
This step matters operationally: it converts transient LLM output into a static artifact that can be used repeatedly in low-resource settings (e.g., Google Forms (e.g., via Apps Script / API tooling) or local LMS import), without requiring the LLM at runtime.
In the AI2L framing, this is one concrete way to implement ``black-box minimization'': the LLM is used for draft generation, while the deployed learning material remains inspectable, editable, and lightweight.

\section{Discussion}
\paragraph{Black-box minimization and self-hosted deployment.}
Our case study suggests that a self-hosted (API-free) LLM can generate deployable MCQs when paired with constrained decoding and deterministic QC gates.
Crucially, L2Q is designed so that learners do \emph{not} need an LLM at practice time: once questions are curated and exported (JSONL/CSV), the learning workflow becomes a conventional quiz session.
We refer to this design stance as \emph{black-box minimization}: a black-box model may assist drafting, but the final deliverable is a static, reviewable artifact and not a system that must keep calling an opaque model.

\paragraph{AI2L as a rubric lens.}
We do not claim a ``direct'' AI2L implementation, because MCQ drafting still relies on a black-box generator.
However, AI2L \cite{shintani2025ai2l} provides a clear lens for communicating which risks are reduced by design and which remain.
Table~\ref{tab:ai2l_lens} summarizes the mapping between the four pillars and our concrete pipeline elements.

\begin{table}[t]
\centering
\small
\caption{AI2L pillars as a rubric lens for this work (alignment and remaining gaps).}
\label{tab:ai2l_lens}
\begin{tabular}{p{0.25\linewidth} p{0.15\linewidth} p{0.15\linewidth} p{0.15\linewidth} p{0.24\linewidth}}
\toprule
\textbf{AI2L pillar} & \textbf{Local inference} & \textbf{QC + rubric} & \textbf{Static export} & \textbf{Remaining gap} \\
\midrule
Transparency (black-box exclusion) & \checkmark & \checkmark & \checkmark & Drafting model remains opaque \\
Accountability (human verification) & --- & \checkmark & \checkmark & Human final approval still required \\
Privacy (info protection) & \checkmark & \checkmark & \checkmark & Depends on local material handling \\
Green AI (sustainability) & \checkmark & --- & \checkmark & Authoring still compute-heavy \\
\bottomrule
\end{tabular}
\end{table}

\paragraph{What QC can guarantee---and what it cannot.}
The QC layer targets \emph{machine-checkable} properties that are necessary for safe deployment: schema compliance, single-answer structure, intra-run de-duplication, and numeric/constant equivalence.
QC can also \emph{flag} clarity risks (e.g., missing rounding instructions), but it cannot guarantee pedagogical validity (objective alignment, difficulty calibration, fairness, or misconception targeting).
Accordingly, we position QC as a \emph{gatekeeper for formal correctness} and as a \emph{review accelerator} for humans, not as a replacement for expert judgment.

\paragraph{Practical extensions.}
Instructors can iterate seeds/prompts until both QC and rubric checks are satisfactory.
The QC checker can be extended with domain-specific constraints (unit consistency, dimensional analysis, prerequisite graphs).
Finally, local deployment reduces exposure of proprietary lecture materials and allows institutions to adopt stricter data-handling policies.

\section{Limitations and Responsible Use}
\paragraph{Scale and external validity.}
Our experiments focus on a small entropy mini-pack (three short ``dummy'' lecture PDFs) and report \emph{pipeline-level} QC metrics rather than educational outcomes.
Accordingly, the present evidence should be interpreted as an engineering validation of reproducibility and failure-mode coverage, not as proof of improved learning.

\paragraph{Long and heterogeneous PDFs.}
L2Q relies on extracting text segments that fit the LLM context window.
For long lecture notes (tens to hundreds of pages), naive chunking can fragment prerequisite chains, blur topic boundaries, and increase duplicates or near-duplicates.
Stable coverage for such documents may require hierarchical indexing (section\,$\rightarrow$\,subsection\,$\rightarrow$\,paragraph), retrieval, and/or multi-pass generation, which increases compute cost and complicates reproducibility.

\paragraph{Figure/table-heavy or scanned documents.}
The current pipeline is text-first and does not perform robust document-vision/OCR understanding.
When key information is conveyed in figures, tables, or equations embedded as images (common in STEM PDFs), text extraction may miss essential evidence.
In these cases the model may ``fill in'' missing details, and deterministic QC gates cannot reliably detect that the underlying source support was absent.

\paragraph{Mathematical equivalence checking is partial.}
Our automated checkers catch several common pitfalls (e.g., duplicated numeric constants, approximate constant equality within tolerance, and sampled equivalence tests), but they are not a complete symbolic prover.
Algebraically equivalent expressions written in different forms may be flagged as different (false negatives), while expressions that coincide only on sampled points or within floating tolerances may be treated as equivalent (false positives).
This limitation becomes more acute for unit-aware quantities, piecewise definitions, and items requiring multi-step derivations.

\paragraph{QC guarantees vs.\ what remains human work.}
QC is best understood as a \emph{risk-reduction layer}, not an automatic truth oracle.

\textbf{QC can reliably guarantee (within our design):}
\begin{itemize}[leftmargin=*]
\item \textbf{Structural validity:} JSON schema conformance and required fields; exactly five options; exactly one marked correct.
\item \textbf{Certain deterministic equivalences:} detection of duplicated numeric/constant distractors (e.g., $\log_2(0.25)$ vs.\ $2\log_2(0.5)$) and uniqueness of the marked-correct option under the same equivalence tests.
\item \textbf{Actionable red flags:} warnings for likely ambiguity patterns such as missing rounding/precision instructions when the correct option is approximate.
\end{itemize}

\textbf{QC cannot automatically guarantee:}
\begin{itemize}[leftmargin=*]
\item \textbf{Semantic correctness and faithfulness:} whether the question truly matches the lecture content, whether explanations are conceptually correct, or whether the distractors are pedagogically meaningful.
\item \textbf{Deep mathematical equivalence:} general symbolic equivalence (beyond the numeric/constant checks we implement) or equivalence under nontrivial transformations.
\item \textbf{Coverage and learning impact:} curricular alignment, difficulty calibration, or educational effectiveness for a target learner population.
\end{itemize}

We therefore recommend treating QC as triage: it reduces the search space for expert review rather than replacing it.

\paragraph{Model and prompt dependence.}
Although we publish prompts, parameters, manifests, and file checksums, outcomes still depend on the chosen LLM, quantization, decoding settings, and toolchain versions.
Seed sweeps quantify variance under a fixed setup but do not remove model dependence; re-running with a different model or context size may change both the warning profile and content quality.

\paragraph{Privacy, copyright, and energy.}
Local inference reduces exposure of proprietary lecture materials and can avoid repeated cloud calls, but it does not by itself solve licensing or data governance.
Practitioners should follow institutional policies, consider energy/compute budgets when scaling to large corpora, and treat all exported items as instructor-reviewed educational content.

\section{Conclusion}
We presented L2Q, a self-hosted (API-free) lecture-to-quiz pipeline that uses a local LLM to draft MCQs, applies deterministic QC gates, and exports static artifacts ready for quiz platforms. Across 15 runs on an entropy mini-pack, the pipeline achieved QC acceptance 1.0 with low retry overhead, producing a deployable set of 24 curated questions. We framed the approach as black-box minimization and discussed its relationship to AI2L principles as a rubric lens. Future work includes broader classroom evaluations, richer pedagogical rubrics, and further reductions of black-box dependence even in the authoring stage.

\section{Data and Code Availability}\label{sec:availability}
All supplementary artifacts are provided as arXiv ancillary files under \texttt{anc/}:
\begin{itemize}[leftmargin=*]
  \item \texttt{anc/arxiv\_dummy\_entropy\_pack.zip}: the three dummy lecture PDFs (inputs) and a minimal end-to-end reproduction notebook (prompts, parameters, and QC rules).
  \item \texttt{anc/seed\_sweep\_outputs.zip}: per-run outputs and QC logs for the 15-run seed sweep, plus an aggregated run summary and per-run \texttt{run\_manifest.json} files.
  \item \texttt{anc/final24\_entropy\_mcq.jsonl} and \texttt{anc/final24\_entropy\_mcq.csv}: the curated 24-question deploy set exported in import-friendly formats.
\end{itemize}
Because the input lectures are synthetic/dummy documents, redistribution does not pose copyright or privacy issues.

\section*{Acknowledgements}
All experiments reported in this paper and the manuscript writing were performed solely by the author.
The author thanks Yuri Amano for her undergraduate thesis presentation and early exploration of self-hosted (API-free) lecture-to-quiz workflows, which helped motivate the present study.
The author also thanks Associate Professors Hiroaki Takadama and Seiji Yamaguchi for valuable discussions in the weekly seminar and for sharing laboratory facilities.
This work was supported by JSPS KAKENHI Grant Number JP25K00269 (Grant-in-Aid for Scientific Research (C), Project Title: ``Elucidation of Myosin Molecular Dynamics Associated with Sarcomere Morphological Changes in the Intracellular Environment''), by Chubu University FY2025 Special Research Fund (CP) (``Development of an AI safety evaluation and AI2L-based utilization system for clinical laboratory technologist education''), and by the Chubu University FY2025 Research Institute for Industry and Economics (RIIE) Research Project (``Development and evaluation of an explainable management-analysis support tool using local small-scale LLMs'').
Any remaining errors are the author's responsibility.

\bibliographystyle{unsrtnat}
\bibliography{references}

\appendix

\section{Artifact packaging}\label{app:bundle}
The accompanying ancillary files (\texttt{anc/}) contain all components needed to reproduce the reported results: (i) three dummy lecture PDFs, (ii) a runnable end-to-end pipeline notebook, (iii) per-run outputs and QC reports for the 15-run seed sweep, and (iv) the curated 24-item deploy set in JSONL/CSV. See Section~\ref{sec:availability} for the exact filenames.

\section{Prompt templates (summary)}
\label{app:prompt}
Below is a condensed version of the question-generation prompt template (placeholders in braces).
The full, runnable templates (generation + verifier) are included in the artifact bundle.

\begin{verbatim}
You are an instructor creating high-quality multiple-choice questions (MCQs)
based ONLY on the lecture excerpt.

OUTPUT FORMAT (STRICT JSON ONLY):
- Output must be a single JSON object only (no markdown, no commentary).
- Create exactly ONE MCQ with 5 choices: OptionA..OptionE.
- CorrectOption must be one of: A, B, C, D, E (single letter).
- SourcePages must be included (e.g., "p.1" or "p.1-2").

{"SourcePages":"p.1-2","Question":"...?",
 "OptionA":"...","OptionB":"...","OptionC":"...","OptionD":"...","OptionE":"...",
 "CorrectOption":"A","Explanation":"...",
 "ID":"Q{q_index:02d}","Category":"MCQ","FocusTopic":"{focus_topic}"}

CONTENT RULES:
- Each question must be answerable using the excerpt.
- Use exactly 5 options; exactly 1 correct.
- Avoid external knowledge; avoid multiple-correct or ambiguous items.
- Keep explanations short and grounded in the excerpt.
- Avoid repeating previously asked questions (a short list is provided in the prompt).

NOTE:
The pipeline repeats this ONE-MCQ call with different topic hints until
n_questions=8 accepted items are collected per run (with bounded retries).
\end{verbatim}

\section{QC rules and logging schema}
\label{app:qc}
Each run emits a machine-readable manifest (config + timing), a QC report (hard failures, retries, warnings), and normalized datasets.
For each MCQ item, we use the following schema:

\begin{table}[h]
\centering
\small
\caption{Internal per-item schema used in run-level artifacts (\texttt{questions.json} / \texttt{QuestionBank.csv}).}
\label{tab:item_schema_internal}
\begin{tabular}{lp{0.68\linewidth}}
\toprule
Field & Description \\
\midrule
\texttt{ID} & Unique ID within the run (e.g., \texttt{Q01}). \\
\texttt{SourcePages} & Page range supporting the item (traceability). \\
\texttt{Question} & MCQ stem. \\
\texttt{OptionA}--\texttt{OptionE} & Five answer choices. \\
\texttt{CorrectOption} & Correct choice letter (A--E). \\
\texttt{Explanation} & Short rationale grounded in the excerpt. \\
\texttt{Category} & Category tag (e.g., \texttt{MCQ}). \\
\texttt{FocusTopic} & Topic-plan label for coverage control. \\
\bottomrule
\end{tabular}
\end{table}

\begin{table}[h]
\centering
\small
\caption{Released export schema for the curated deploy set (\texttt{final24\_entropy\_mcq.jsonl} / \texttt{final24\_entropy\_mcq.csv}).}
\label{tab:item_schema_release}
\begin{tabular}{lp{0.68\linewidth}}
\toprule
Field & Description \\
\midrule
\texttt{lecture} & Lecture identifier (e.g., \texttt{Lecture01}). \\
\texttt{run} & Selected run tag (e.g., \texttt{RUN4}). \\
\texttt{seed} & RNG seed for the run. \\
\texttt{pdf\_basename} & Input PDF filename (basename). \\
\texttt{source\_pages} & Page pointer supporting the item. \\
\texttt{id} & Stable item ID (e.g., \texttt{L01\_Q01}). \\
\texttt{category} & Category tag. \\
\texttt{focus\_topic} & Topic-plan label. \\
\texttt{question} & MCQ stem. \\
\texttt{options} & (JSONL) map \texttt{\{"A":..., ... "E":...\}}. \\
\texttt{option\_A}--\texttt{option\_E} & (CSV) flattened option columns. \\
\texttt{correct} & Correct choice letter (A--E). \\
\texttt{explanation} & Short rationale. \\
\bottomrule
\end{tabular}
\end{table}

In addition, each run directory includes \texttt{run\_manifest.json} capturing the run configuration, timing, and file checksums (and, where available, tool/version identifiers). Hard QC failures trigger re-generation (bounded by \texttt{retry\_max}). Warnings (e.g., Rnd, Dup) are logged to facilitate efficient human review.

\section{Representative MCQ items from the final 24-question set}
\label{app:sample_mcqs}
To make the outputs concrete, we reproduce six representative MCQs (two per lecture) from the curated 24-item deploy set.
We select items with distinct pedagogical ``roles'' so readers can quickly see coverage: (1) definition/concept check, (2) numeric calculation, (3) conceptual law statement, (4) formula-based computation, (5) microstate-to-entropy interpretation, and (6) canonical probability weighting.
Minor typesetting changes (e.g., writing $\log_2$ in math mode) are applied for readability; the semantic content matches the released JSONL/CSV.

\subsection*{Lecture01: Information theory (entropy in bits)}
\noindent\textbf{\texttt{L01\_Q05}} (SourcePages: p.1--3; FocusTopic: Chain rule for joint entropy; Role: concept/property check)\\
\textbf{Question.} According to the chain rule in information theory, what is the relationship between the joint entropy $H(X,Y)$ and the conditional entropy $H(Y\mid X)$?\\
\textbf{A.} $H(X,Y)=H(X)-H(Y\mid X)$\\
\textbf{B.} $H(X,Y)=H(X)+H(Y\mid X)$\\
\textbf{C.} $H(X,Y)=H(X)\cdot H(Y\mid X)$\\
\textbf{D.} $H(X,Y)=H(X)\,/\,H(Y\mid X)$\\
\textbf{E.} $H(X,Y)=H(X\mid Y)-H(Y\mid X)$\\
\textbf{Correct.} B\\
\textbf{Explanation.} The chain rule states $H(X,Y)=H(X)+H(Y\mid X)$.\\[0.8em]

\noindent\textbf{\texttt{L01\_Q07}} (SourcePages: p.1--3; FocusTopic: Worked example --- entropy of a biased coin; Role: numeric calculation)\\
\textbf{Question.} For a biased coin with $p(1)=0.9$ and $p(0)=0.1$, what is the approximate Shannon entropy $H(X)$ in bits?\\
\textbf{A.} 0.1368 bits\\
\textbf{B.} 0.3322 bits\\
\textbf{C.} 0.469 bits\\
\textbf{D.} 0.6 bits\\
\textbf{E.} 1 bit\\
\textbf{Correct.} C\\
\textbf{Explanation.} $H(X)=-0.9\log_2(0.9)-0.1\log_2(0.1)\approx 0.469$ bits.\\[1.0em]

\subsection*{Lecture02: Thermodynamics (second law and entropy)}
\noindent\textbf{\texttt{L02\_Q02}} (SourcePages: p.1--3; FocusTopic: Clausius statement of the second law; Role: conceptual law statement)\\
\textbf{Question.} Which statement best describes the Clausius statement of the second law of thermodynamics?\\
\textbf{A.} The entropy of an isolated system always decreases over time.\\
\textbf{B.} Heat cannot spontaneously flow from a colder object to a hotter object.\\
\textbf{C.} The total energy of an isolated system is not conserved.\\
\textbf{D.} It is possible to convert all heat into work in a cyclic process.\\
\textbf{E.} The entropy of the surroundings always decreases when a system expands.\\
\textbf{Correct.} B\\
\textbf{Explanation.} The Clausius statement says heat does not spontaneously transfer from cold to hot.\\[1.0em]

\noindent\textbf{\texttt{L02\_Q07}} (SourcePages: p.1--3; FocusTopic: Entropy change for reversible isothermal expansion; Role: formula-based computation)\\
\textbf{Question.} In a reversible isothermal expansion of an ideal gas, which expression gives the entropy change $\Delta S$?\\
\textbf{A.} $\Delta S=nR\ln(T_2/T_1)$\\
\textbf{B.} $\Delta S=nR\ln(V_2/V_1)$\\
\textbf{C.} $\Delta S=nC_V\ln(T_2/T_1)$\\
\textbf{D.} $\Delta S=Q/T$ for any irreversible process\\
\textbf{E.} $\Delta S=0$ for all expansions\\
\textbf{Correct.} B\\
\textbf{Explanation.} For reversible isothermal expansion, $\Delta S=nR\ln(V_2/V_1)$.\\[1.0em]

\subsection*{Lecture03: Statistical mechanics (microstates, ensembles)}
\noindent\textbf{\texttt{L03\_Q02}} (SourcePages: p.1--3; FocusTopic: Boltzmann entropy formula; Role: microstate-to-entropy link)\\
\textbf{Question.} What is the formula for entropy in statistical mechanics according to Boltzmann?\\
\textbf{A.} $S=k_B\Omega$\\
\textbf{B.} $S=k_B\ln\Omega$\\
\textbf{C.} $S=\ln(k_B\Omega)$\\
\textbf{D.} $S=\Omega/k_B$\\
\textbf{E.} $S=k_B/\ln\Omega$\\
\textbf{Correct.} B\\
\textbf{Explanation.} Boltzmann's formula is $S=k_B\ln\Omega$, where $\Omega$ is the number of microstates.\\[1.0em]

\noindent\textbf{\texttt{L03\_Q05}} (SourcePages: p.1--3; FocusTopic: Canonical ensemble probability; Role: canonical probability weighting)\\
\textbf{Question.} In the canonical ensemble, what is the probability $p_i$ of a system being in microstate $i$ with energy $E_i$?\\
\textbf{A.} $p_i=E_i/Z$\\
\textbf{B.} $p_i=1/\Omega$\\
\textbf{C.} $p_i=\exp(-\beta E_i)/Z$\\
\textbf{D.} $p_i=Z\,\exp(\beta E_i)$\\
\textbf{E.} $p_i=Z/\exp(\beta E_i)$\\
\textbf{Correct.} C\\
\textbf{Explanation.} In the canonical ensemble, $p_i=\exp(-\beta E_i)/Z$, where $\beta=1/(k_BT)$ and $Z$ is the partition function.

\end{document}